# Anomalous Superconductivity in Twisted $MoTe_2$ Nanojunctions

Yanyu Jia,[1*†] Tiancheng Song,[1†] Zhaoyi Joy Zheng,[1,2] Guangming Cheng,[3] Ayelet J Uzan,[1] Guo Yu,[1,2] Yue Tang,[1] Connor J. Pollak,[4] Fang Yuan,[4] Michael Onyszczak,[1] Kenji Watanabe,[5] Takashi Taniguchi,[6] Shiming Lei,[7] Nan Yao,[3] Leslie M Schoop,[4] N. P. Ong,[1] Sanfeng Wu[1*]

[1] Department of Physics, Princeton University, Princeton, New Jersey 08544, USA.
[2] Department of Electrical and Computer Engineering, Princeton University, Princeton, New Jersey 08544, USA.
[3] Princeton Materials Institute, Princeton University, Princeton, New Jersey 08544, USA.
[4] Department of Chemistry, Princeton University, Princeton, New Jersey 08544, USA.
[5] Research Center for Electronic and Optical Materials, National Institute for Materials Science, 1-1 Namiki, Tsukuba 305-0044, Japan.
[6] Research Center for Materials Nanoarchitectonics, National Institute for Materials Science, 1-1 Namiki, Tsukuba 305-0044, Japan.
[7] Department of Physics, The Hong Kong University of Science and Technology, Clear Water Bay, Kowloon 999077, Hong Kong, China.
† These authors contributed equally to this work
*Email: sanfengw@princeton.edu; yanyuj@princeton.edu

**Abstract**

**Introducing superconductivity in topological materials can lead to innovative electronic phases and device functionalities. Here, we present a new strategy for quantum engineering of superconducting junctions in moiré materials through direct, on-chip, and fully encapsulated 2D crystal growth. We achieve robust and designable superconductivity in Pd-metalized twisted bilayer molybdenum ditelluride ($MoTe_2$) and observe anomalous superconducting effects in high-quality junctions across ~ 20 moiré cells. Surprisingly, the junction develops enhanced, instead of weakened, superconducting behaviors, exhibiting fluctuations to a higher critical magnetic field compared to its adjacent $Pd_7MoTe_2$ superconductor. Additionally, the critical current further exhibits a striking V-shaped minimum at zero magnetic field. These features are unexpected in conventional Josephson junctions and indeed absent in junctions of natural bilayer $MoTe_2$ created using the same approach. We discuss implications of these observations, including the possible formation of mixed even- and odd-parity superconductivity at the moiré junctions. Our results also demonstrate a pathway to engineer and investigate superconductivity in fractional Chern insulators.**

## Introduction

Recent observations of fractional quantum anomalous Hall effect(*1–3*) in twisted bilayer molybdenum ditelluride (t$MoTe_2$)(*4–6*) have confirmed the existence of fractional Chern insulators (FCI)(*7–13*) in the absence of magnetic fields. The creation of superconductivity (SC) in FCIs can in principle lead to new electronic states of matter under novel experimental conditions.(*14*, *15*) However, it is challenging to create SC using traditional means in such air-sensitive two-dimensional (2D) moiré materials. Here we overcome such challenges by presenting a new strategy for constructing high-quality superconducting junctions consisting of air-sensitive van der Waals(vdW) moiré materials, such as $tMoTe_2$. We present systematic characterizations of the junction, including both the atomic structure and the electronic transport behaviors.

## Results

### vdW-Encapsulated 2D Growth and Nanojunctions

The key to our approach of fabricating high-quality superconducting moiré junctions is the recently introduced on-chip 2D growth mechanism (*16*), which is based on the surprising discovery of a rapid mass transport and crystal growth templated on 2D materials. The air-sensitive 2H-$MoTe_2$ flakes, in contact with pre-deposited palladium (Pd) source, are together fully encapsulated between top and bottom graphite/hexagonal boron nitride (hBN) stacks. The Pd serves as the seed of growth in the next step (see **Figs. 1A & B** for cartoon illustrations of the device and crystal structure). The device as fabricated contains no superconducting materials (neither Pd nor $MoTe_2$ superconducts). We then anneal the device at ~ 185 ºC, which triggers the transport of an ultrathin uniform layer of Pd into $MoTe_2$ layer and their reactions produce a new crystalline compound $Pd_7MoTe_2$, which can be seen as the darker region extending from the Pd contacts in the optical image of the final device after annealing (**Fig. 1C**). All processes involving $MoTe_2$ are implemented in an Argon-filled glovebox to prevent degradation (Method section). Detailed characterizations of such a chamber-free 2D low-temperature synthesis, generalizable to various combinations of metals and 2D materials, can be found in ref(*16*). Interestingly, using this approach we find that a class of new Pd-based compounds produced on topological chalcogenides, including the $Pd_7MoTe_2$ synthesized here, are superconductors(*17*).

Here, we demonstrate a controllable creation of high-quality superconducting junctions in bilayer $MoTe_2$ in both its natural and twisted forms. **Figures 1D-G** show a typical growth of $Pd_7MoTe_2$ on $tMoTe_2$ in device D1 in which we record the synthesis process under an atomic force microscope after selected growth elapsed time. The length of the $tMoTe_2$ junction between two $Pd_7MoTe_2$ islands can be accurately determined by controlling the growth time. The increased thickness in the vdW stack due to the Pd spread is ~ 1.5 nm (**Fig. 1H**), consistent with our previous report(*16*).

### Atomic Characterization of the Junction

We first characterize the new compound $Pd_7MoTe_2$ and the nanojunction using scanning transmission electron microscope (S/TEM), following the fabrication and experimental procedures described in refs (*16*, *18*) for both cross-section and plan-view STEM studies. **Fig. 2A** shows a STEM image of a suspended $tMoTe_2$ film placed on a Pd-deposited TEM grid (sample T1), after the heat treatment at 190 ºC for 6 minutes. The lateral growth of the new compound (brighter area) from the Pd outer seed leads to a junction of $Pd_7MoTe_2$/$tMoTe_2$/ $Pd_7MoTe_2$ where the $tMoTe_2$ gap is about 30 nm wide. We performed energy-dispersive X-ray spectroscopy (EDX) analysis on the film (**Fig. 2B**) and confirmed that the atomic ratio of Pd/Mo/Te in the new compound is always very close to 7:1:2. **Fig. S1** shows EDX analysis on other locations that are well separated, confirming the uniformity of the as-grown material. **Fig. 2C** displays high-angle annular dark field (HAADF) images with an atomic resolution, which clearly reveals both a crystalline structure of the $Pd_7MoTe_2$ and the moiré lattice of $tMoTe_2$. The fast Fourier transformation (FFT) pattern of the region $tMoTe_2$ (**Fig. 2D**) confirms its lattice structure as well as a twist angle of 3.7º, which is the target angle during fabrication. The FFT pattern of $Pd_7MoTe_2$, as shown in **Fig. 2E**, is indistinguishable from our previous observation (*16*) of $Pd_7WTe_2$, implying that the two crystal structures are the same. It is remarkable that an exceptionally sharp lateral interface (< 1 nm) between $Pd_7MoTe_2$ and $tMoTe_2$ is achieved and that the moiré structure of $tMoTe_2$ remains intact in the junction (**Fig. 2F**). **Fig. S2** shows the cross-section STEM images taken from a $Pd_7MoTe_2$/$tMoTe_2$ device (sample T2) grown inside a vdW stack fully encapsulated by hBN, like the transport device shown in **Fig. 1**. The data there again confirm the high quality of the junction.

We note that our gentle and low-temperature growth helps improve the moiré homogeneity in the $tMoTe_2$ region because the ultrathin Pd film serves as a glue that seals the two $MoTe_2$ monolayers, an advantage of our approach.

**Anomalous Superconductivity in $tMoTe_2$ Junctions**

In the transport study of this work, we focus on SC properties of junctions with a width $d \sim 100$ nm, which we call short junctions. We demonstrate that SC is indeed achieved across the junction of $tMoTe_2$ in device D2, fabricated with an interlayer twist angle of $\sim 3.7^{\circ}$ and a junction $d \sim 105$ nm (across ~ 20 moiré cells) (**Fig. 3A & B**). **Figure 3A** illustrates our transport measurement scheme that detects the voltage drops both on the $Pd_7MoTe_2$ and across the $tMoTe_2$ junction when current passes through. The measured resistances ($R_{xx}$) *v.s.* temperature ($T$) exhibit sharp decreases to zero just below ~ 1 K for both $Pd_7MoTe_2$ superconductor (*17*) and the $tMoTe_2$ junction (**Fig. 3C**). The $IV$ characteristics across the junction (**Fig. 3D)** display the expected SC nonlinearity and a sharp transition at a critical current ($I_c$) of ~ 42 nA. The differential resistance $dV/dI$ *v.s.* applied DC current ($I$), taken for both $Pd_7MoTe_2$ and the junction, reveals a smaller critical current across the junction (**Fig. 3E**).

However, we observe major differences between the moiré junctions and a conventional Josephson junction. The first hint of anomaly is that the Sc fluctuations of the junction exist up to a higher temperature than $Pd_7MoTe_2$. As shown in **Fig. 3C**, $T_c$, chosen as the temperature at which $R_{xx}$ reaches half of the normal state value, of the junction (~ 0.88 K) is higher than that of the $Pd_7MoTe_2$ superconductor (~ 0.78 K). We further highlight this feature in **Figs. 3F & G**, where $dV/dI$ curves are recorded upon warming up the device. When $Pd_7MoTe_2$ is in its normal state ($T > 0.9$ K), substantial nonlinearity in $IV$ curves remains across the junction at ~1.5 K.

The most striking features of the moiré junctions emerge when a magnetic field ($B$) is applied normal to the film. **Figures 4A-D** display the critical current behaviors of both $Pd_7MoTe_2$ and the moiré junction as a function of $B$. Whereas SC of $Pd_7MoTe_2$ at 50 mK is fully suppressed above ~ 1.2 T (**Fig. 4A**), features of SC at the junction persist to much higher fields (~ 2.2 T). The contrast is most apparent in **Fig. 4E**, where we compare their resistive transitions vs. $B$. Both higher $T_c$ and higher critical $B$ ($B_c$) suggests that the pairing potential in the junction is enhanced over that in the $Pd_7MoTe_2$ pads and that the usual proximity effects are not sufficient to explain our data. The data suggest that superconductivity with a pairing potential distinct from that in $Pd_7MoTe_2$ appears to reside in the junction.

This is accompanied by another unexpected feature in the field profile of the critical current. In a conventional Josephson junction, winding of the supercurrent phase results in a pattern in which $I_c$ peaks at $B = 0$. However, the moiré junctions exhibit a clear "V-shaped" minimum at zero $B$, as seen in **Figs. 4B, C & F**. A different perspective of the anomalous minimum in $I_c$ is seen if we plot d$V$/d$I$ *v.s.* $B$ with $I$ fixed at 43 nA. We see that $B$ drives the junction from a resistive state (at $B$=0) to a dissipationless SC state as $B$ increases (**Fig. 4G**). These anomalies, including the enhanced critical $B$ and the V-shaped critical current minimum, are robust under repeated thermal cycling and highly reproducible across devices (see **Figs. S3-S5** for data taken in another device, D3) and at all gate voltages (**Figs. S6 & S7**).

A robust V-shaped minimum in $I_c$ at zero $B$ is quite rare. A well-known example is the corner junction formed between a *d*-wave cuprate superconductor and an s-wave superconductor. In this situation, the destructive interference occurs between two spatially separated supercurrents with a relative phase shift of π.(*19*) Likewise, destructive supercurrent interference can also be observed between spatially separated 0- and π- junctions in superconductor/ferromagnet/superconductor (SFS) junctions.(*20*) In both cases, unconventional pairing and interference effects are key ingredients. We note an important distinction between our moiré junctions and SFS junctions. To see a V-shaped minimum in SFS junctions, one has to fine-tune the junction length $d$ to sub-nanometer accuracy because the exchange splitting energy causes the sign of the Josephson coupling to oscillate rapidly with a spatial period of ~ 1 nm.(*20*) By contrast, the V-shaped minimum here is observed in all devices without fine-tuning $d$ (e.g., $d$ ~ 105 nm in D2 and ~ 90 nm in D3). In addition, the time reversal symmetry is preserved in our junction. Guided by the corner-junction experiment on cuprate superconductors, we reason that the presence of destructive interference at the junction is needed for explaining the observed V-shaped minimum. Yet the situation is also distinct from the cuprate corner junction since in our case the device is better described as a single junction without a corner geometry.

One possibility to reconcile all the experimental facts is to attribute such a destructive interference effect to the coexistence of an odd- and an even-parity condensate in the moiré junction. Both time reversal and inversion symmetries play key roles in the electron pairing of SC. In the presence of both symmetries, the conventional Bardeen–Cooper–Schrieffer theory favors an even-parity spin-singlet pairing. However, in the absence of either symmetry, unconventional pairing may occur. Particularly in materials with strong spin-orbit coupling (SOC) and broken inversion symmetry, SC with mixed even- and odd-parity states are anticipated(*21–23*). The possibility of unconventional pairing in non-centrosymmetric materials, e.g., heavy fermion systems, has been explored in the past decades(*22–24*), although challenges in confirming an odd-parity superconductor remain. In our devices, the spin-orbit coupling (SOC) is quite large in $tMoTe_2$ and the moiré lattice lacks inversion symmetry. Theoretically, this can lead to admixtures of even and odd-parity pair condensates(*21–26*) in the moiré junction. We hence speculate that such a mixed pairing condensate may be responsible for generating two supercurrent channels with a π phase shift that destructively interfere. Whether the two channels are spatially overlapping or spontaneously separated awaits further experimental tests.

**Unusual Normal State Conduction in the Junction**

We further remark on the normal state resistance observed in short junctions ($d$ ~100 nm). $MoTe_2$ is an insulator with a large activation gap ($\Delta$ ~ 1 eV). However, in short junctions we observe metallic behavior (even without applying a gate voltage) with a normal state resistance of several kiloohms that is nearly $T$-independent up to room temperature (**Fig. 3C**, **Figs. S3** & **S8**). The normal state resistance also exhibits little gate dependence (**Fig. S8**). Below $T_c$, these metallic-like junctions exhibit the supercurrents as described. We note that (1) the atomic resolution STEM studies of the junction (**Fig. 2** and **Fig. S2**) confirm the absence of Pd atoms in the $tMoTe_2$ region in the junction and the moiré structure remains unchanged, and (2) conduction due to tunneling events between rare hoping site, due to e.g., disorders, cannot explain the metallicity of the junction, since such tunneling conductance should strongly depend on temperature, in contrast to the experimental observations (**Fig. S8A**).

One possible explanation of the unusual residual conduction is to assume that the band-bending effects close to the metallic $Pd_7MoTe_2$ pads could cause conduction channels in $tMoTe_2$ junction tobe populated, although the junction width of ~ 100 nm seems quite large for such an effect. The induced electron density needed in the junction seems also quite high since we cannot deplete it with electrostatic gating. Note that, for longer junctions with $d > 500$ nm, our experiments do show that the gate-tuned insulator state of tMoTe2 is recovered, following expectations. We currently do not have a comprehensive explanation considering all the experimental facts. Future experiments, such as low temperature scanning tunneling microscopy, would help resolve the situation.

Although the exact mechanisms for the normal state conduction and superconductivity remain to be worked out, we next show that (1) the metallicity in these short junctions occur for both $tMoTe_2$ and natural bilayer $MoTe_2$ and yet (2) only the $tMoTe_2$ junction exhibits the superconducting anomalies. Namely, both the V-shaped minimum and the enhanced pair potential vanish if we perform the same experiments on the inversion symmetric natural bilayer.

**Contrasting Behaviors in a Natural Bilayer Junction**

We now repeat the experiments described above on natural bilayer $2H$-$MoTe_2$, which hosts an inversion center located in the middle of the two layers (**Figs. 5A & B**). The junction is fabricated using the same approach, with a similar length $d$ ~ 100 nm. Both the unusual normal state conduction and the SC are consistently observed in the natural bilayer junction, yet the anomalous supercurrent features seen in $tMoTe_2$ are now absent. **Figs. 5C-E** display the $T$-dependent resistances of both the natural bilayer junction and the $Pd_7MoTe_2$ in the same device, similar to previous discussions. Characteristics of the $Pd_7MoTe_2$ superconducting pads in this device (D4) are closely similar to those in the moiré devices (D2 & D3), with similar values for the normal state resistance. Values of $I_c$ when measured across the junctions are also similar. $T_c$ of the junction is now slightly lower than the $Pd_7MoTe_2$ SC (**Fig. 5C**). The SC anomalies are now absent in the natural bilayer junction. (1) $B_c$ of the junction also no longer exceeds that of $Pd_7MoTe_2$ (both ~ 1.2 T) (**Figs. 5F & G**). Crucially, (2) the V-shaped minimum in the $I_c$ at zero $B$ is now replaced by the conventional maximum as seen in **Figs. 5 G-I**. The absence of the SC anomalies is confirmed at all gate voltages in the natural bilayer device (**Fig. S9**). We conclude that key properties of the moiré junction, especially the absence of inversion symmetry, are responsible for the superconducting anomalies.

## Discussion

Future experimental and theoretical studies are necessary to uncover the underlying physics. While the understanding of the SC pairing symmetry certainly requires future experiments, the approach here based on moiré materials suggests a promising strategy to study unconventional pairing in non-centrosymmetric superconductors. The sharp interface between $Pd_7MoTe_2$ and $tMoTe_2$ also implies a new route for engineering SC in moiré topological materials with the goal of proximitizing fractionalized states. Our current devices have a twist angle of ~ 3.7°, which has been shown to host the integer and fractional Chern insulator states upon electrostatic gating(*1*, *2*, *4*–*6*). Preliminary results in the current devices reveal a weak but interesting gate modulation of the junction critical current (**Fig. S10**). Investigating the coexistence of SC and FCI states is possible with further optimization of the devices.

## Materials and Methods

### Device fabrication

**Transport Devices (D1-D4).** hBN and graphite flakes were exfoliated on $SiO_2$/Si substrates, identified and characterized under optical microscopes and atomic force microscopes (AFM, Bruker Dimension Edge or Bruker Dimension Icon). Subsequently, hBN flakes were stacked on top of graphite flakes and then placed on $SiO_2$/Si substrates. Electron beam lithography, followedby cold development, reactive ion etching and metal deposition, were employed to create Pd contacts and growth seeds (~ 20 nm thick) on the bottom hBN/graphite stack. Prior to final assembly, the bottom stacks were then tip cleaned using AFM under the contact mode. To prepare the top stacks, we exfoliated monolayers and bilayers 2H-$MoTe_2$ in an Ar-filled glovebox. For $tMoTe_2$ devices (D1, D2 and D3), monolayer $MoTe_2$ was cut into 2 pieces using a sharp tungsten tip. The first piece was picked up by the top vdW stack consisting of hBN and graphite flakes. The second piece of $MoTe_2$ underwent a 3.7° rotation before being stacked with the first piece. For device D4, natural bilayer $MoTe_2$ was directly picked up by a top hBN/graphite stack. The top stacks of $MoTe_2$/hBN/graphite, for both types of devices, were then carefully aligned and positioned on the prepared bottom stacks. The devices prepared above were then AFM tip cleaned before being placed on a hot plate for the on-chip growth of $Pd_7MoTe_2$. With controlled temperature and time, the Pd growth process was carefully monitored under an optical microscope and atomic force microscope. To achieve the precise control, devices are initially monitored under optical microscope until the two $Pd_7MoTe_2$ pads are approximately 500 nm apart. Then, according to the estimated growth rate, the devices are examined under AFM after each short-time extra growth to achieve the targeted junction length. The entire process involving $MoTe_2$ was performed in a glovebox filled with argon, with concentration of $H_2O$ < 0.1 ppm and $O_2$ < 0.1 ppm.

**TEM Devices (T1 and T2).** The suspended TEM device (T1) was fabricated by stacking a 3.7° $tMoTe_2$ onto a Pd-coated TEM grid using standard dry transfer technique. The polycarbonate (PC) used for the dry transfer was later removed by dissolving in chloroform for 30 mins. Pd was introduced into the $tMoTe_2$ by holding the TEM grid at 190 °C for 6 minutes inside the vacuum chamber of STEM. For the TEM cross-section device (T2), the $tMoTe_2$ stack and $Pd_7MoTe_2$ growth was created using the same process as for transport devices. A lamella specimen was then extracted from a selected region of the stack using a standard lift-out technique within a focused ion beam-scanning electron microscope (FIB-SEM) system. The specimen was further thinned and polished using a $Ga^+$ ion beam until it became sufficiently transparent for STEM analysis. All fabrication steps of both devices, including the removal of the PC layer, were conducted in an Ar-filled glovebox with $H_2O$ < 0.1 ppm and $O_2$ < 0.1 ppm. More fabrication details can be found in (*16, 18*)

**Transport measurements**

The electrical transport measurement was conducted in a dilution refrigerator equipped with a superconducting magnet and a base temperature of ~ 20 mK. Four-probe resistance measurements were performed using the standard ac lock-in technique with a low frequency, typically ~ 23.3 Hz, and an ac current excitation from 0.5 nA to 10 nA. Additionally, a dc current is also applied for critical current measurements.

**STEM measurements**

Atomic resolution high-angle annular dark-field (HAADF) STEM imaging and energy dispersive X-ray spectroscopy (EDX) mappings were performed on a Titan Cubed Themis 300 double Cs-corrected scanning/transmission electron microscope (S/TEM), equipped with an extreme field emission gun source and a super-X EDS system. The system was operated at 300 kV. A Gatan double tilt heating holder (Model 652) was used for in-situ heating study.

**Supplementary Materials**

This PDF file includes:
Figs. S1 to S10

## Acknowledgments

We acknowledge discussions with Xiaodong Xu, Biao Lian and B. Andrei Bernevig, and thank Xiaodong Xu and Jiaqi Cai for sharing their bulk $MoTe_2$ crystals.

**Funding:** This work is mainly supported by AFOSR Young Investigator Award (FA9550-23-1-0140) to S.W. Electric transport measurement is partially supported by NSF through the Materials Research Science and Engineering Center (MRSEC) program of the National Science Foundation (DMR-2011750) through support to L. M. S. and S.W. and a CAREER award (DMR-1942942) to S.W. Device fabrication is partially supported by ONR through a Young Investigator Award (N00014-21-1-2804) to S.W. N.P.O. acknowledges support from the United States Department of Energy (DE-SC0017863) and the Gordon and Betty Moore Foundation through Grants GBMF9466. S.W. and L.M.S. acknowledge support from the Eric and Wendy Schmidt Transformative Technology Fund at Princeton. S.W. acknowledges support from the Gordon and Betty Moore Foundation through Grants GBMF11946 and the Sloan Foundation. L.M.S. acknowledges support from the Gordon and Betty Moore Foundation through Grants GBMF9064 and the David and Lucile Packard Foundation. Y.J. acknowledges support from the Princeton Charlotte Elizabeth Procter Fellowship program. T.S. acknowledges support from the Princeton Physics Dicke Fellowship program. A.J.U acknowledges support from the Rothschild Foundation and the Zuckerman Foundation. C.J.P. is supported by the NSF Graduate Research Fellowship Program under grant number DGE-2039656. K.W. and T.T. acknowledge support from the JSPS KAKENHI (Grant Numbers 21H05233 and 23H02052) and World Premier International Research Center Initiative (WPI), MEXT, Japan. S.L. acknowledge the financial support provided by the start-up fund of the Hong Kong University of Science and Technology, and the Hong Kong Collaborative Research Fund (No. C6053-23G). The authors acknowledge the use of Princeton's Imaging and Analysis Center (IAC), which is partially supported by the Princeton Center for Complex Materials (PCCM), a NSF Materials Research Science and Engineering Center (DMR-2011750).

**Author contributions:** S.W. and Y.J. conceived and designed the project. Y.J. and T.S. fabricated and characterized the transport devices and performed measurements, assisted by G.Y., A. J. U., Y.T., M.O., and Z. J. Z. Y.J. and Z. J. Z. fabricated TEM samples. Y.J., Z. J. Z., G.C. and N.Y. performed STEM measurements assisted by F.Y. C. P., S. L. and L.M.S. grew bulk $MoTe_2$ crystals. K.W. and T.T. provided hBN crystals. S.W., Y. J., and N.P.O. analyzed the data, interpreted the results, and wrote the paper with input from all authors. S.W. supervised the project.

**Competing interests:** Authors declare that they have no competing interests.

**Data and materials availability:** All data needed to evaluate the conclusions are presented in the paper and/or the Supplementary Materials. Additional data related to this paper are available from the corresponding author upon reasonable request.

## Figures

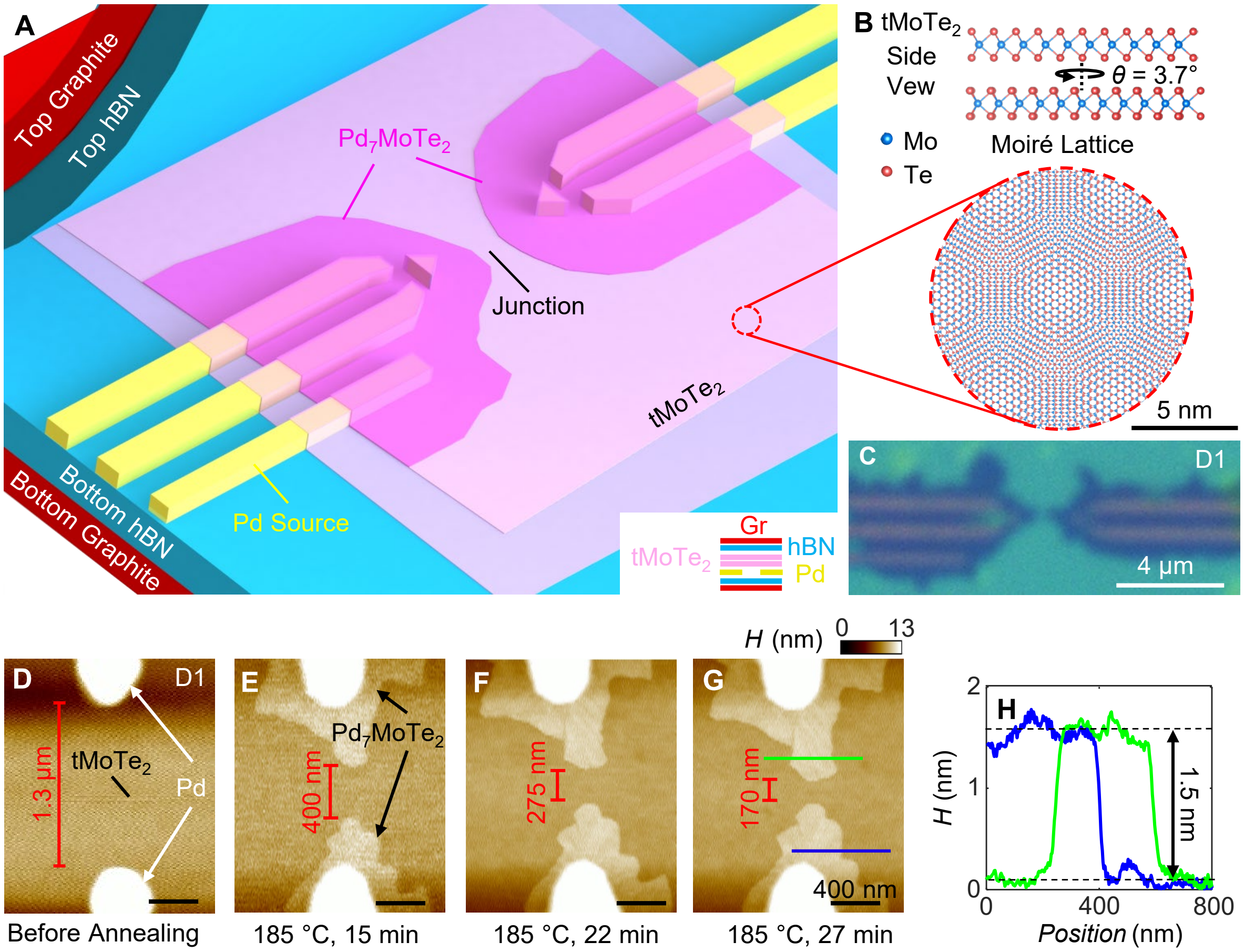

**Fig. 1. vdW-encapsulated, on-chip 2D growth on moiré materials and $tMoTe_2$ junctions. A,** Cartoon illustration of the growth of $Pd_7MoTe_2$ superconductors and the device design. The cross-section device structure is illustrated in the inset at the bottom right. The junction including $tMoTe_2$, $Pd_7MoTe_2$, as well as the Pd seed are fully encapsulated by the top and bottom graphite/ hBN stacks. **B,** $tMoTe_2$ crystal lattice at an interlayer twist angle of ~ 3.7°. **C,** Optical image of a typical device after on-chip growth**.** The darker regions extended from Pd contacts are thin $Pd_7MoTe_2$. **D-G**, A series of atomic force microscope (AFM) images of junction taken during a typical growth process with temperature and elapsed times are indicated below the images. The length of the $tMoTe_2$ junctions is indicated by the red texts. **H,** AFM height profiles along the solid lines indicated in **G** with matched color.

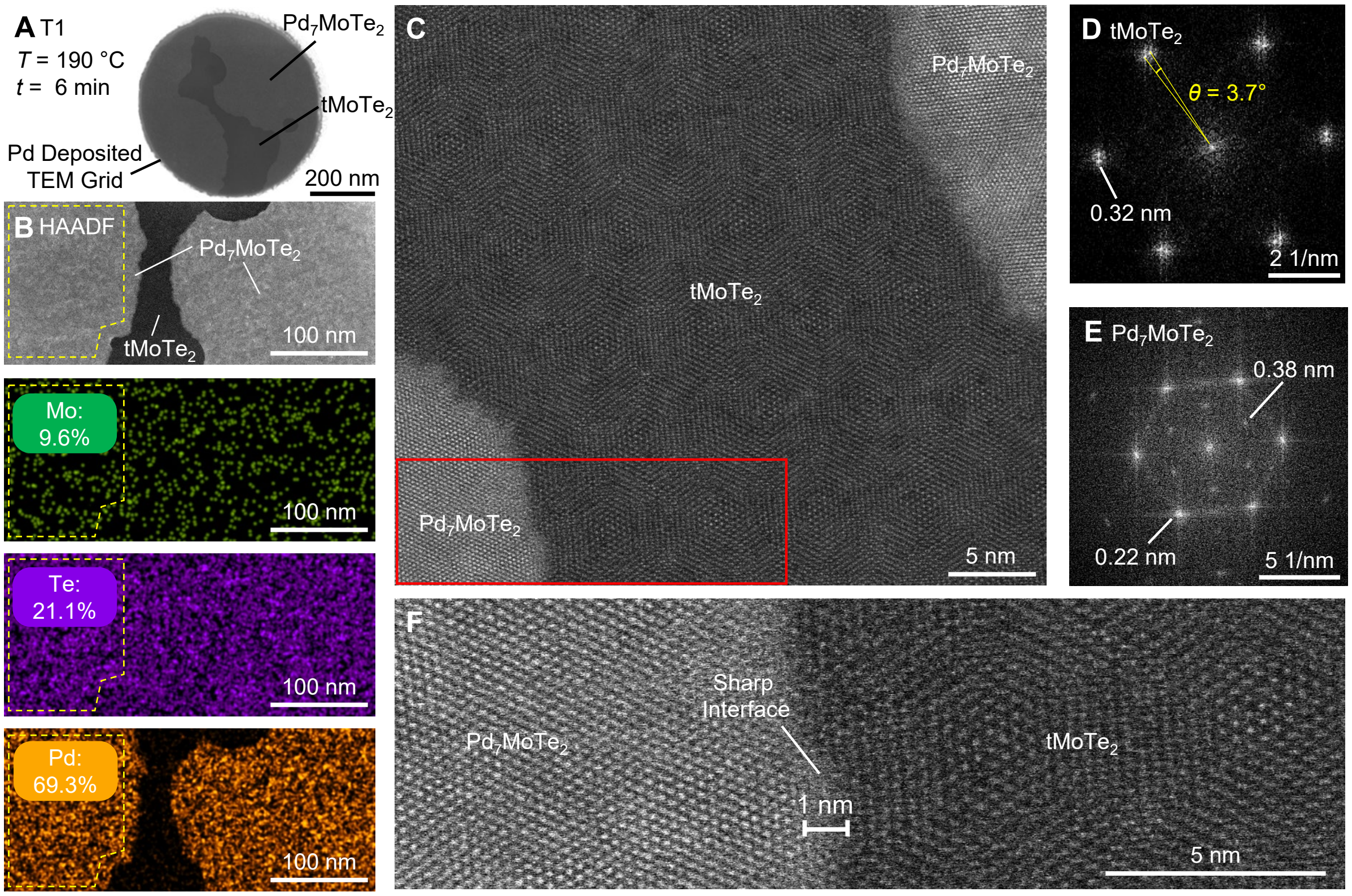

**Fig. 2. STEM analysis of $tMoTe_2$ moiré junction with sharp interfaces. A,** STEM image of $tMoTe_2$ after the growth of $Pd_7MoTe_2$, prepared on a TEM grid (sample T1). The TEM grid is pre-deposited with Pd, followed by transferring $tMoTe_2$ on top. Subsequently, Pd is introduced by holding the temperature at 190°C for 6 minutes, resulting in a $tMoTe_2$ moiré junction with a width of approximately 30 nm. Regions of different materials are indicated. **B,** HAADF and corresponding elemental mappings captured at the moiré junction. The atomic ratio of Pd:Mo:Te is found to be close to 7:1:2 in the new compound, while negligible Pd can be observed in the neighboring $tMoTe_2$ area. **C,** An atomic-resolution STEM image of the moiré junction. **D,** The FFT pattern of $tMoTe_2$ regime, confirming the twist angle of 3.7°. **E,** The FFT pattern of the crystalline $Pd_7MoTe_2$, showing the six-fold symmetry. **F,** Magnification of the STEM image at the interface marked in the red rectangle in **C**. The crystalline structures of both $Pd_7MoTe_2$ and $tMoTe_2$ can be clearly visualized, demonstrating a sharp interface between the two regions. The moiré structure of $tMoTe_2$ remains intact.

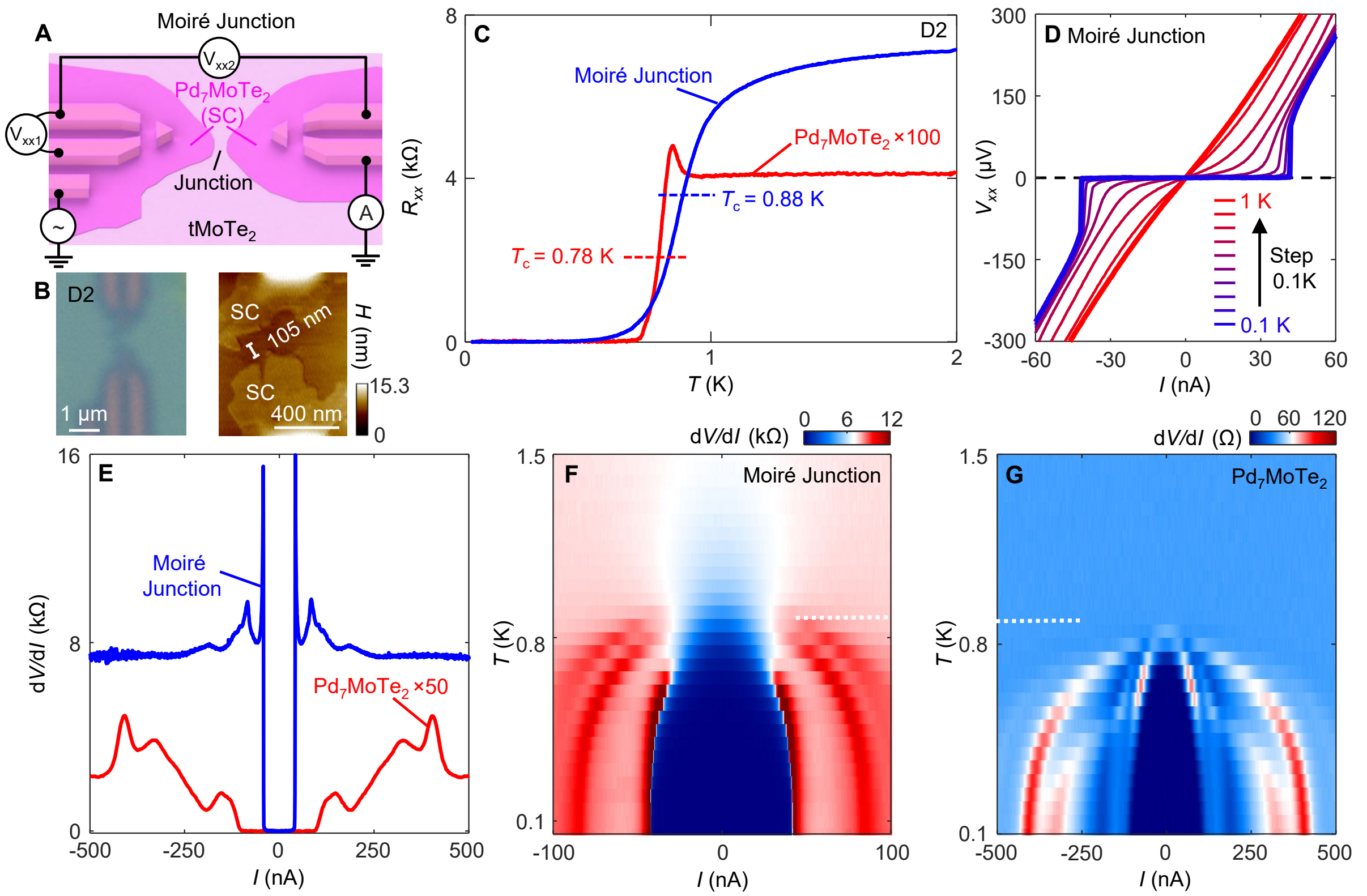

**Fig. 3. Superconductivity across the tMoTe₂ moiré junction. A,** Cartoon illustration of transport measurement scheme. $V_{xx1}$ and $V_{xx2}$ are voltage drops recorded on $Pd_7MoTe_2$ and across the moiré junction, respectively. **B,** Optical (left) and AFM (right) images of D2 (moiré junction, twist angle ~ 3.7°). **C,** Resistance as a function of *T*, for both $Pd_7MoT_2$ (red) and the junction (blue). **D,** *IV* characteristic curves for the moiré junction, taken at various *T* as indicated. **E**, Differential resistance (d*V*/d*I*) *v.s.* an applied DC current (*I*) to the source, for both $Pd_7MoT_2$ (red) and the junction (blue). **F,** d*V*/d*I* *v.s.* *I* taken across the moiré junction under varying *T*. **G,** The same d*V*/d*I* map for $Pd_7MoT_2$ superconductor.

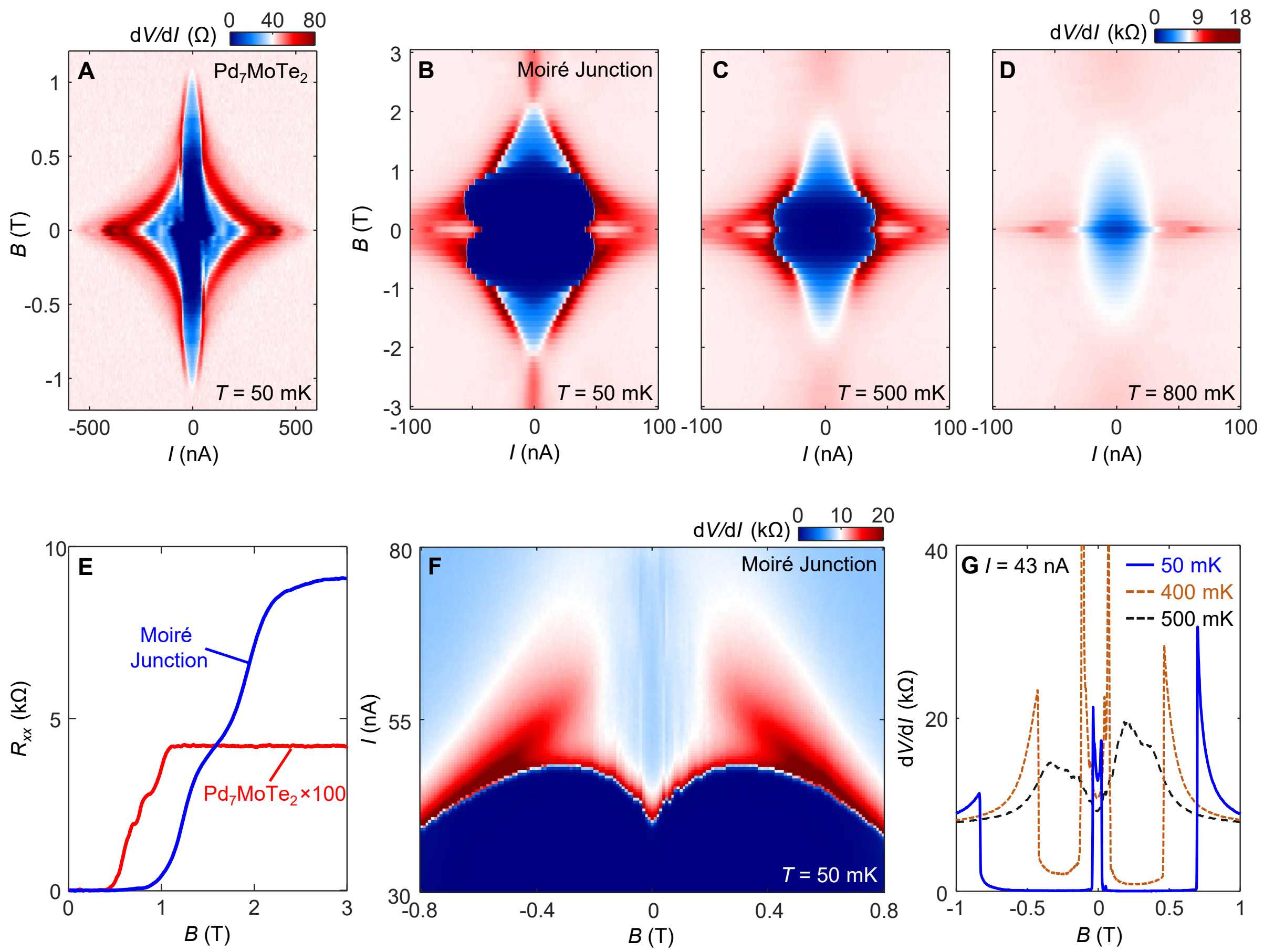

**Fig. 4. Anomalies of the superconducting moiré junction. A,** d$V$/d$I$ v.s. $I$ of the $Pd_7MoTe_2$ superconductor under varying magnetic field ($B$), taken at $T$ = 50 mK. **B,** The same map but for the moiré junction, at $T$ = 50 mK. **C & D,** The same map of the junction at $T$ = 500 mK (**C**) and 800 mK (**D**), respectively. **E,** Resistance as a function of $B$, for both $Pd_7MoTe_2$ (red) and the junction (blue). **F** d$V$/d$I$ map taken under varying $B$ and $I$, at 50 mK, highlighting the V-shaped critical current minimum at zero $B$. **G**, d$V$/d$I$ *v.s.* $B$, taken at a fixed DC current of 43 nA, under three different $T$ as indicated.

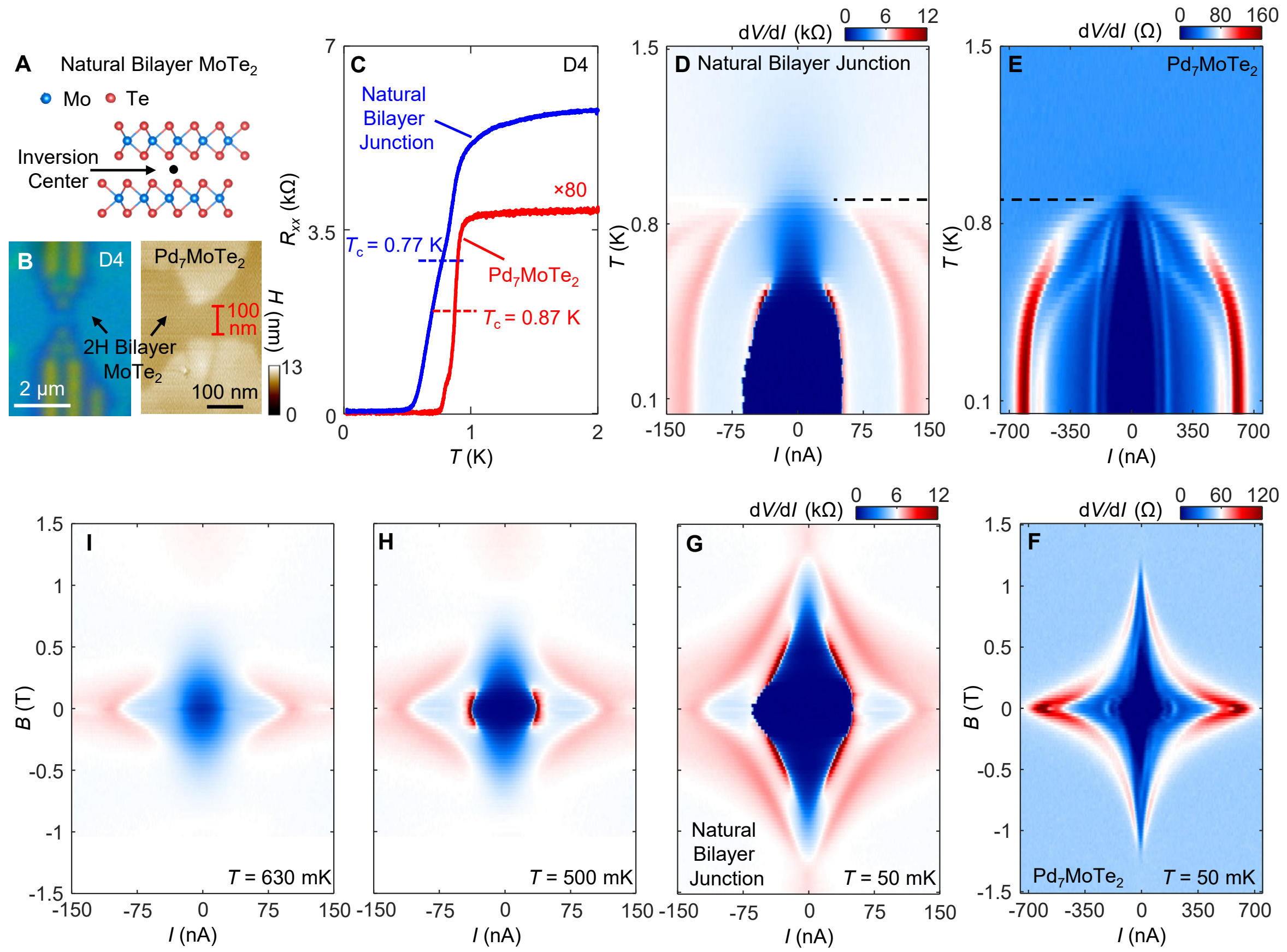

**Fig. 5. Absence of anomalies in the inversion symmetric bilayer junction. A,** Cartoon illustration of exfoliated natural bilayer 2H-$MoTe_2$ lattice structure, where the inversion center is indicated. **B,** An optical image (left) of a natural bilayer device after growth and a corresponding AFM image of the junction (right). **C,** Resistance *v.s.* $T$, taken for $Pd_7MoTe_2$ (red) and the natural bilayer junction (blue) in this device. The same measurement geometry as Fig. 3A is used. **D,** d$V$/d$I$ map of the junction taken under varying $I$ and $T$, showing the SC state. **E,** The same map but for the $Pd_7MoTe_2$ SC. **F,** d$V$/d$I$ map of $Pd_7MoTe_2$, taken under varying $I$ and $B$. **G-I,** The same d$V$/d$I$ maps as **F** but taken for the junction, at three different $T$ as indicated in the maps.

# Supplementary Materials for

## Anomalous Superconductivity in Twisted $MoTe_2$ Nanojunctions

Yanyu Jia et al.

*Corresponding author. Email: sanfengw@princeton.edu; yanyuj@princeton.edu

**This PDF file includes:**

Figs. S1 to S10

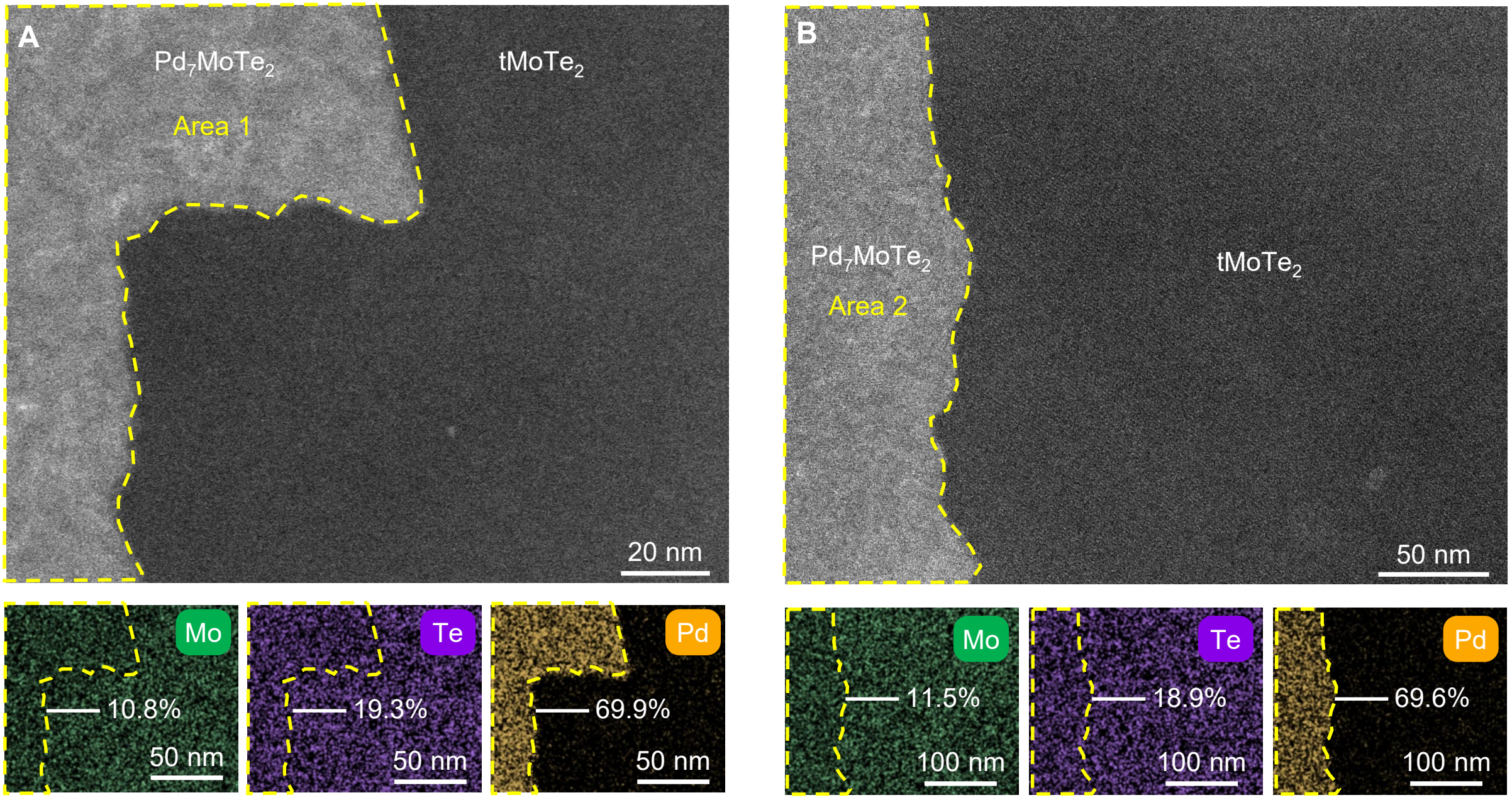

**Fig. S1. Additional EDX analyses and elemental mappings (sample T1). A & B,** The HAADF-STEM images (top) and corresponding EDX elemental mappings (bottom) captured at the interface between $tMoTe_2$ and $Pd_7MoTe_2$ at two additional well-separated locations of T1, show consistent results. The atomic ratios of Pd:Mo:Te in the new compound (outlined by the yellow dashed line) are always found to be approximately 7:1:2.

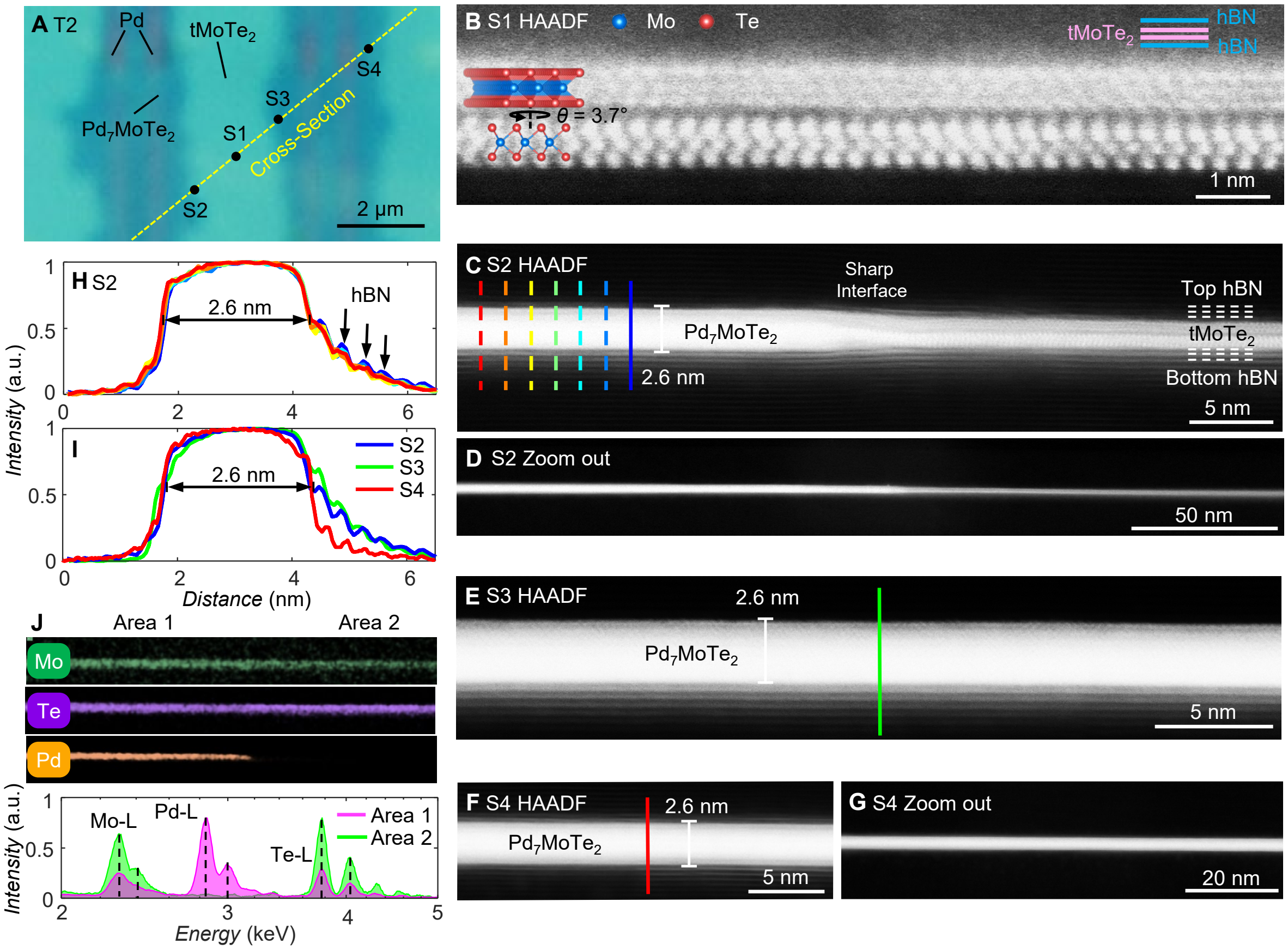

**Fig. S2. Cross-section STEM studies of $Pd_7MoTe_2$ grown on $tMoTe_2$. A,** Optical image of targeted region in sample T2, before being extracted for cross-section STEM studies. The STEM experiment was conducted to visualize the cross-section along the indicated yellow dashed line. Spots S1- S4 are estimated locations under examination in **B-G**. **B,** HAADF-STEM cross-section image of $tMoTe_2$ near spot S1. The atomic structure of bottom $MoTe_2$ layer is clearly resolved and the top layer appears blurry, reflecting the twist. **C,** HAADF-STEM cross-section image captured near S2, located at the boundary between $Pd_7MoTe_2$ and $tMoTe_2$. The thickness of $Pd_7MoTe_2$ is measured to be 2.6 nm. The extracted intensity along the colored line-cuts is shown in **H**. **D,** Zoomed-out STEM image near S2, illustrating the uniform thickness of $Pd_7MoTe_2$. **E-G,** The HAADF-STEM cross-section images taken near spots S3 and S4, located a few microns apart, confirming the same thickness and hence the uniformity of $Pd_7MoTe_2$. **H,** The extracted intensity along the line-cuts in **C**, indicated by the vertical lines. The signal from hBN layer structure is indicated by arrows. **I,** The extracted intensity from S2-S4 along line-cuts indicated by the solid lines in **C, E & F**, again verifying the uniform thickness of $Pd_7MoTe_2$ across the device. **J,** EDX elemental mappings and characteristic X-ray energy spectrum at the boundary between $tMoTe_2$ and $Pd_7MoTe_2$. Pd is only present in the $Pd_7MoTe_2$ area.

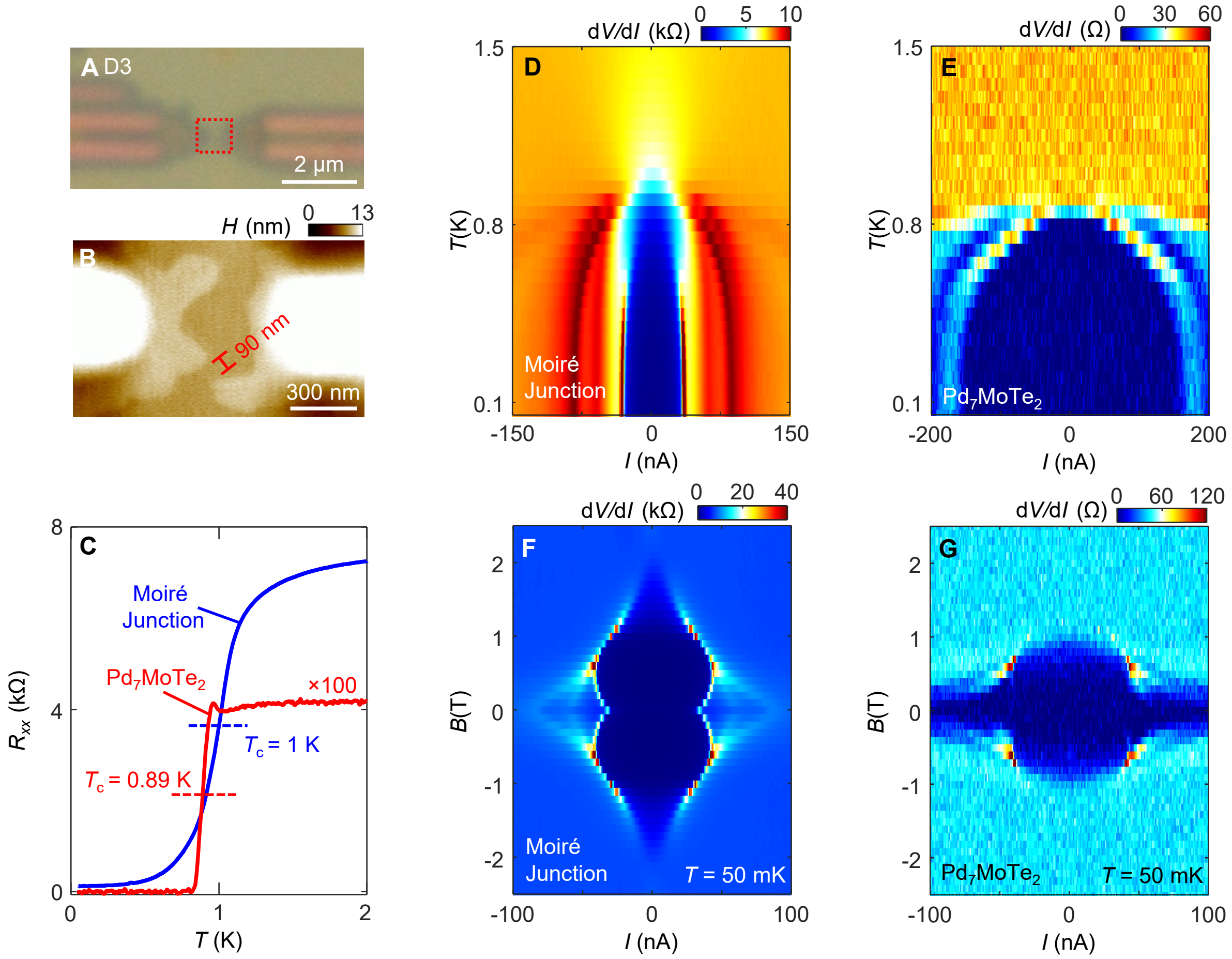

**Fig. S3. Anomalous SC in another $tMoTe_2$ moiré junction (D3). A,** Optical image of device D3 with a $tMoTe_2$ moiré junction (~ 3.7° twist angle) after Pd growth. **B,** AFM image of the moiré junction, revealing a junction length ~ 90 nm at the closest distance. The position of junction is highlighted in red box in **A**. **C,** Resistance *v.s.* $T$, taken for $Pd_7MoTe_2$ (red) and the moiré junction (blue) in this device. The same measurement geometry as Fig. 3A is used. **D & E,** Temperature dependence of $dV/dI$ *v.s.* $I$, taken across the moiré junction (**D**) and on $Pd_7MoTe_2$ (**E**), respectively. **F & G,** Magnetic field ($B$) dependence of $dV/dI$ *v.s.* $I$, taken across the moiré junction (**F**) and on $Pd_7MoTe_2$ (**G**) respectively. Enhanced $T_c$ and $B_c$ and a V-shaped critical current minimum at zero $B$ are robust, reproducible features.

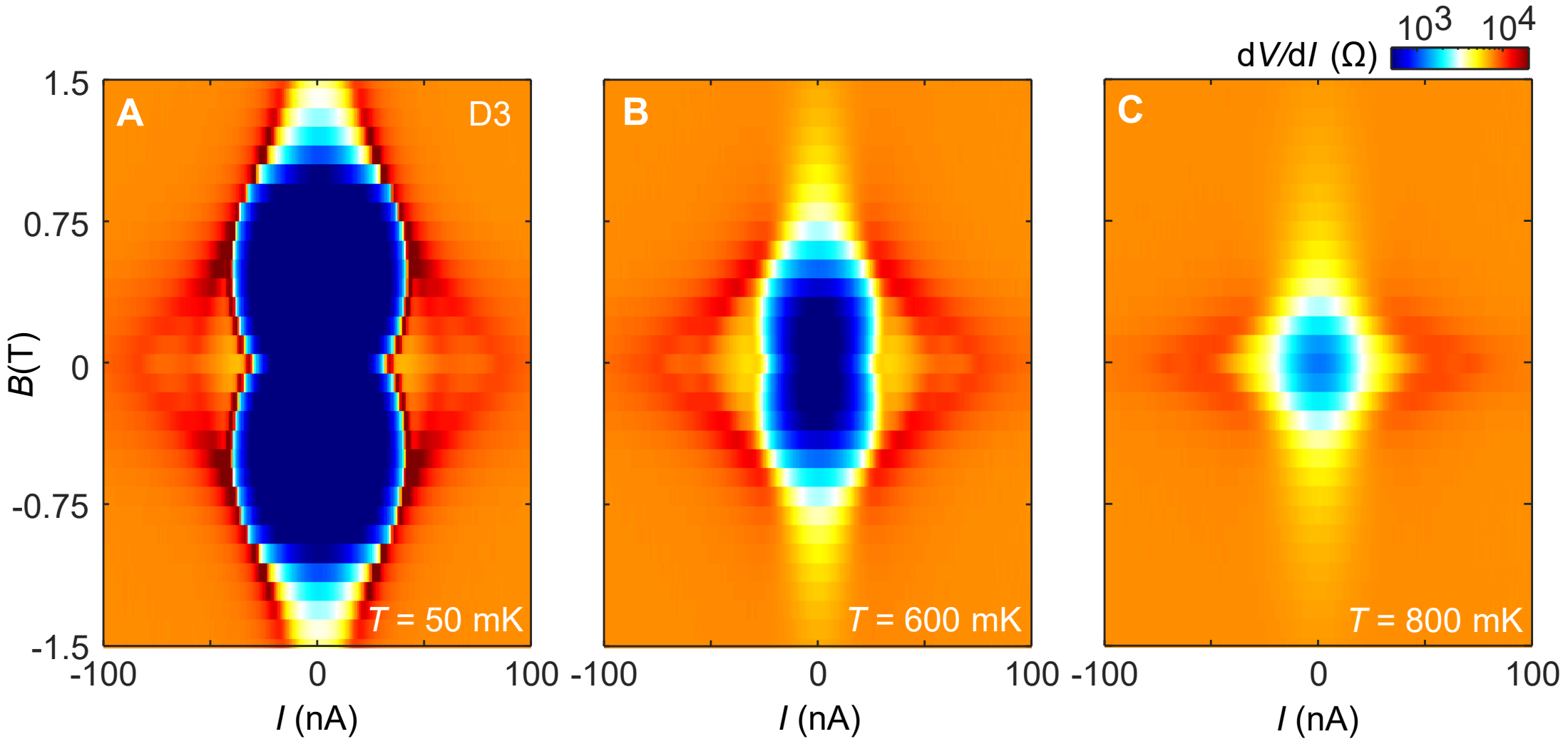

**Fig. S4. Temperature dependence of the V-shaped critical current minimum in D3. A,** The V-shaped critical current minimum observed in D3 at $T = 50$ mK. **B & C,** The same maps but taken at higher temperatures, $T = 600$ mK (**B**) and 800 mK (**C**), respectively.

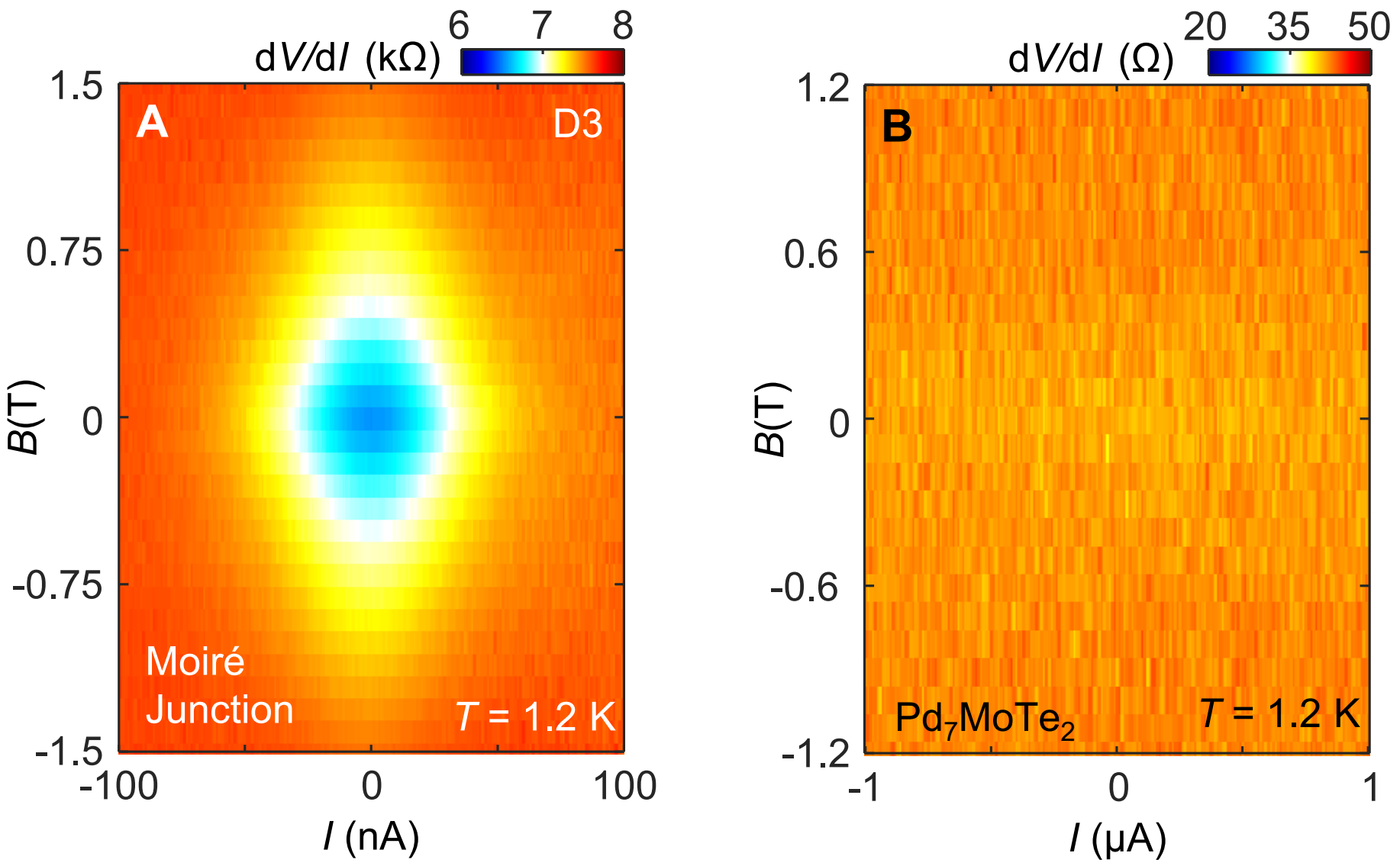

**Fig. S5. Superconducting fluctuations at the moiré junction. A,** d$V$/d$I$ map of the moiré junction, as a function of $B$ and applied DC current $I$, at $T$ = 1.2 K, showing strong superconducting fluctuations. **B,** The same d$V$/d$I$ map for $Pd_7MoTe_2$ taken in the same device, confirming that $Pd_7MoTe_2$ superconductivity is fully suppressed at this $T$.

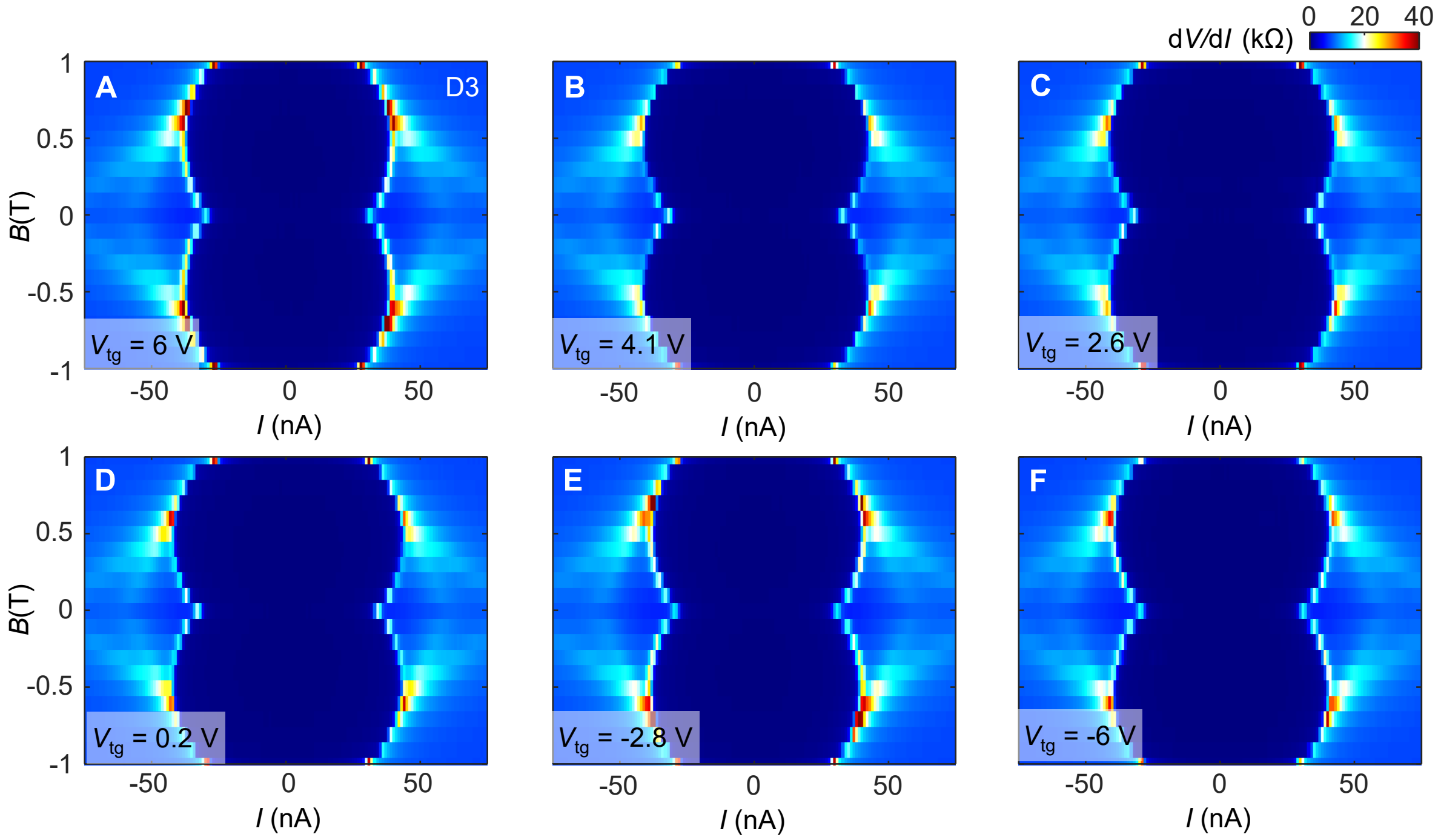

**Fig. S6. Robust V-shaped critical current minimum taken at different gate voltages in D3. A-F,** d$V$/d$I$($B$, $I$) maps of the moiré junction in D3 under varying gates. The V-shaped critical current minimal is observed at all gate voltages (it is also robust against thermocycles).

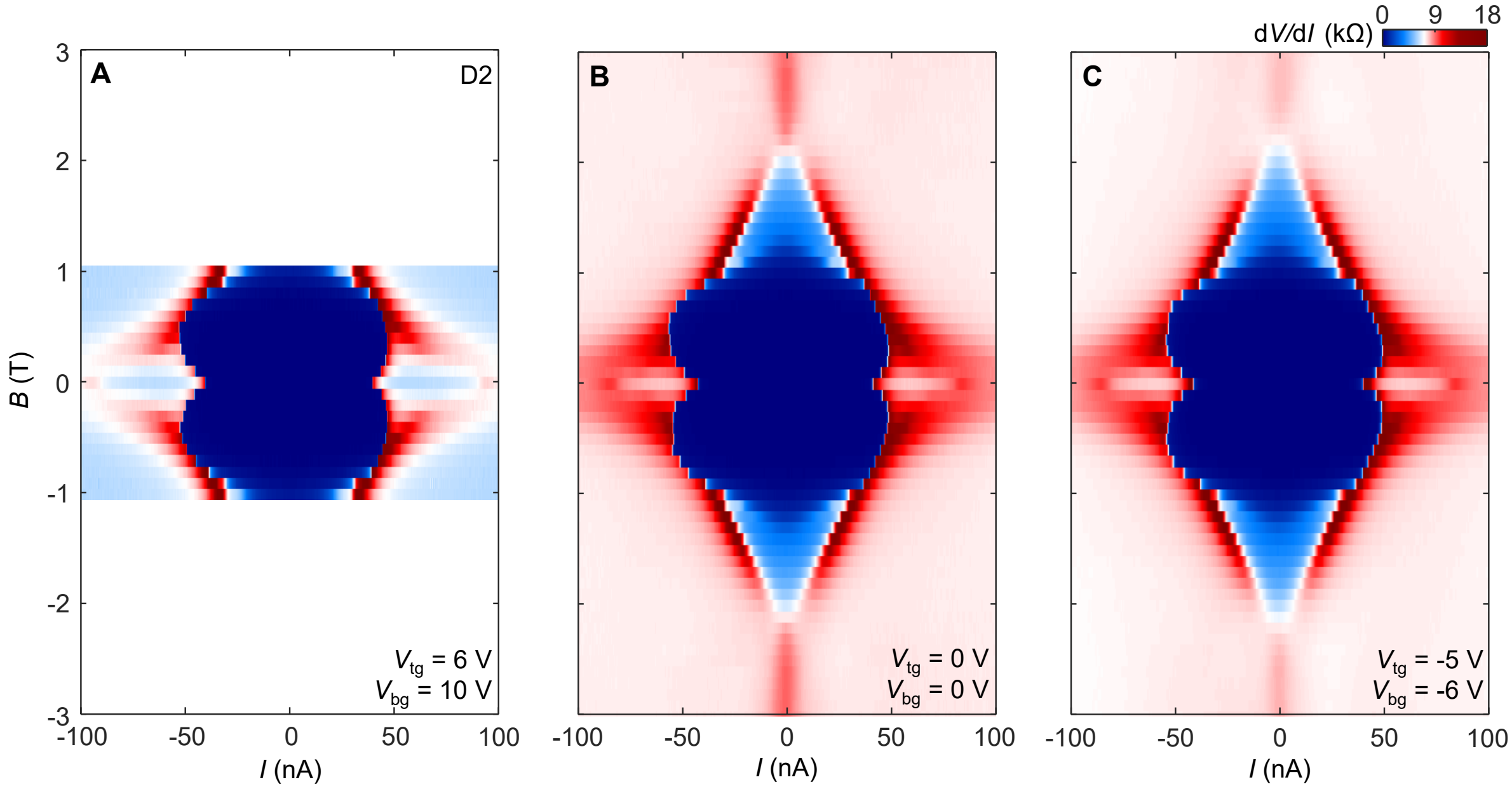

**Fig. S7. Robust V-shaped critical current minimum taken at different gate voltages in D2.** **A-C,** d$V$/d$I$($B$, $I$) maps of the moiré junction in D2 under varying gate voltages and all displays the similar V-shaped critical current minimum at $B$ = 0 T.

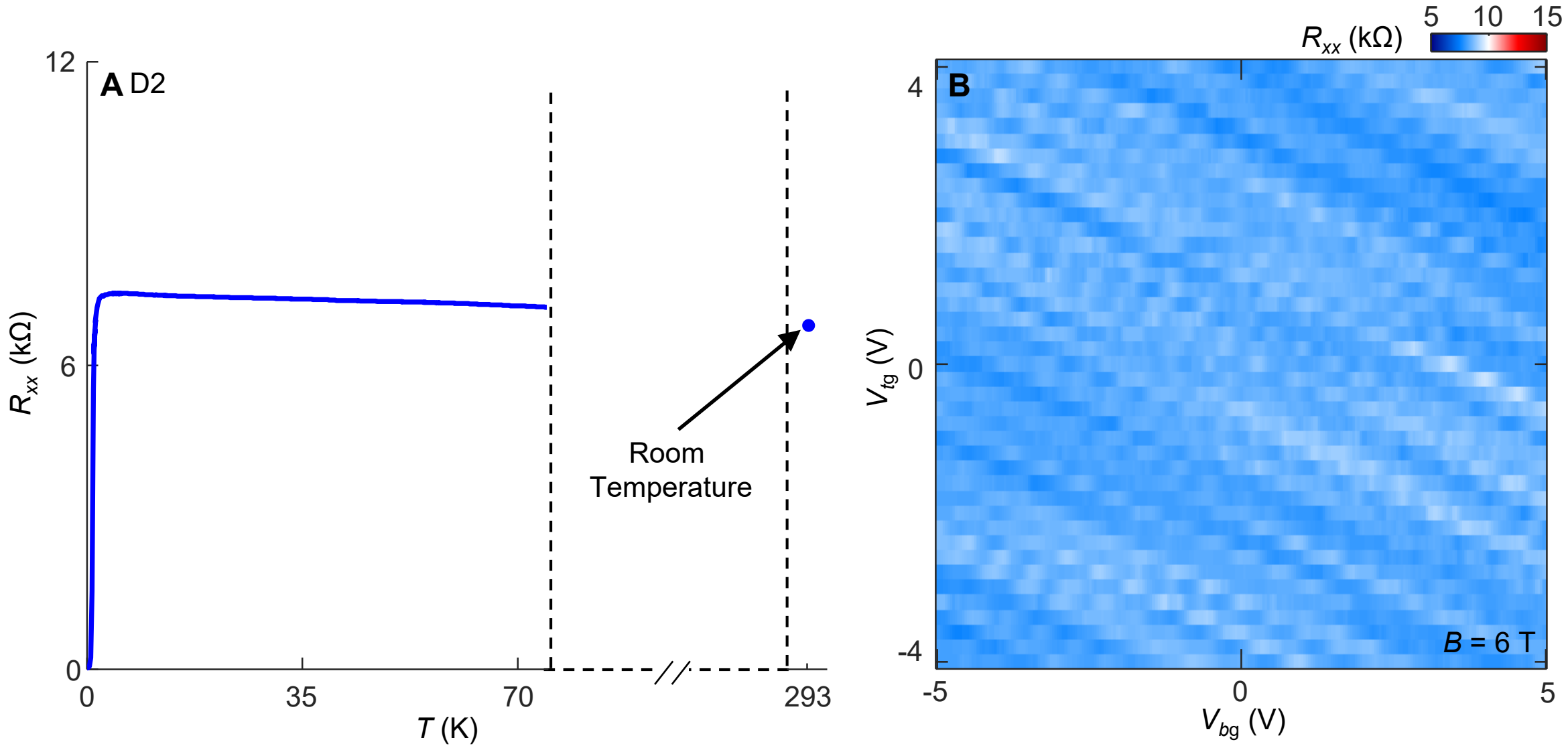

**Fig. S8. Temperature- and gate-dependent normal state resistance ($R_{xx}$) of the moiré junction (D2).** Normal state resistance of the moiré junction shows little temperature (**A**) or gate (**B**) dependence.

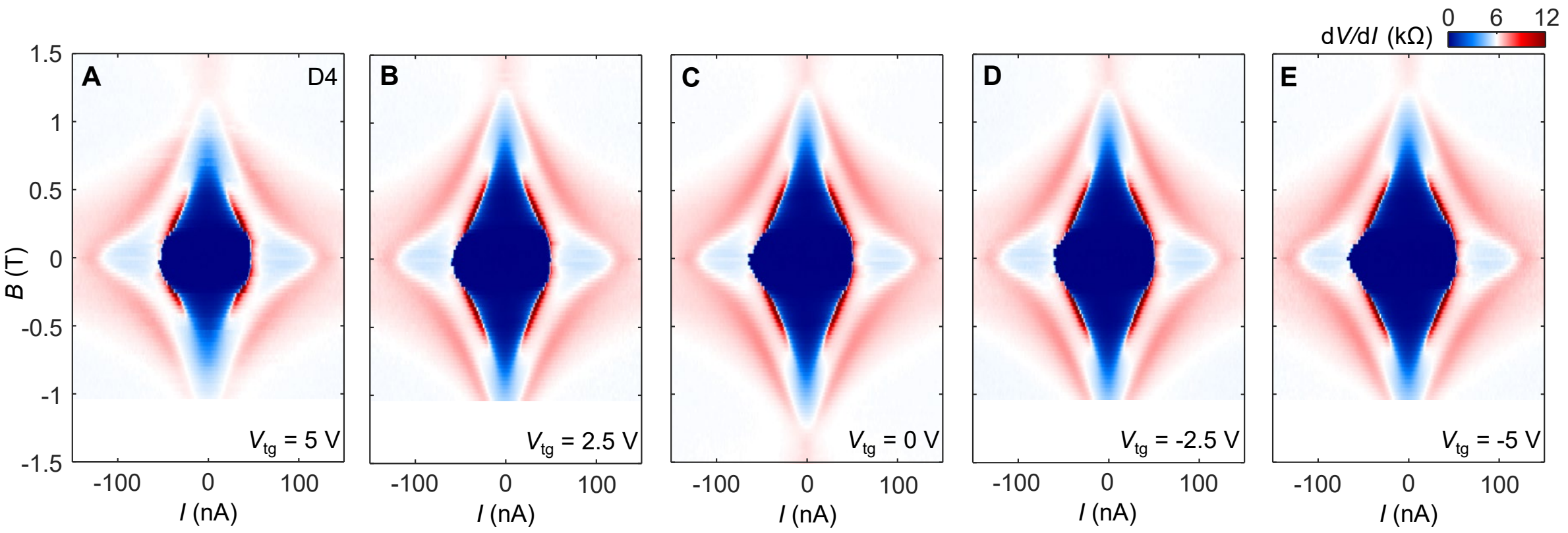

**Fig. S9. The absence of the SC anomalies in natural bilayer junction at all gate voltages (D4). A-E,** Magnetic field ($B$) dependence of d$V$/d$I$ *v.s.* $I$ maps taken across the junction at various gate voltages, all confirming the absence of SC anomalies in the natural bilayer junction.

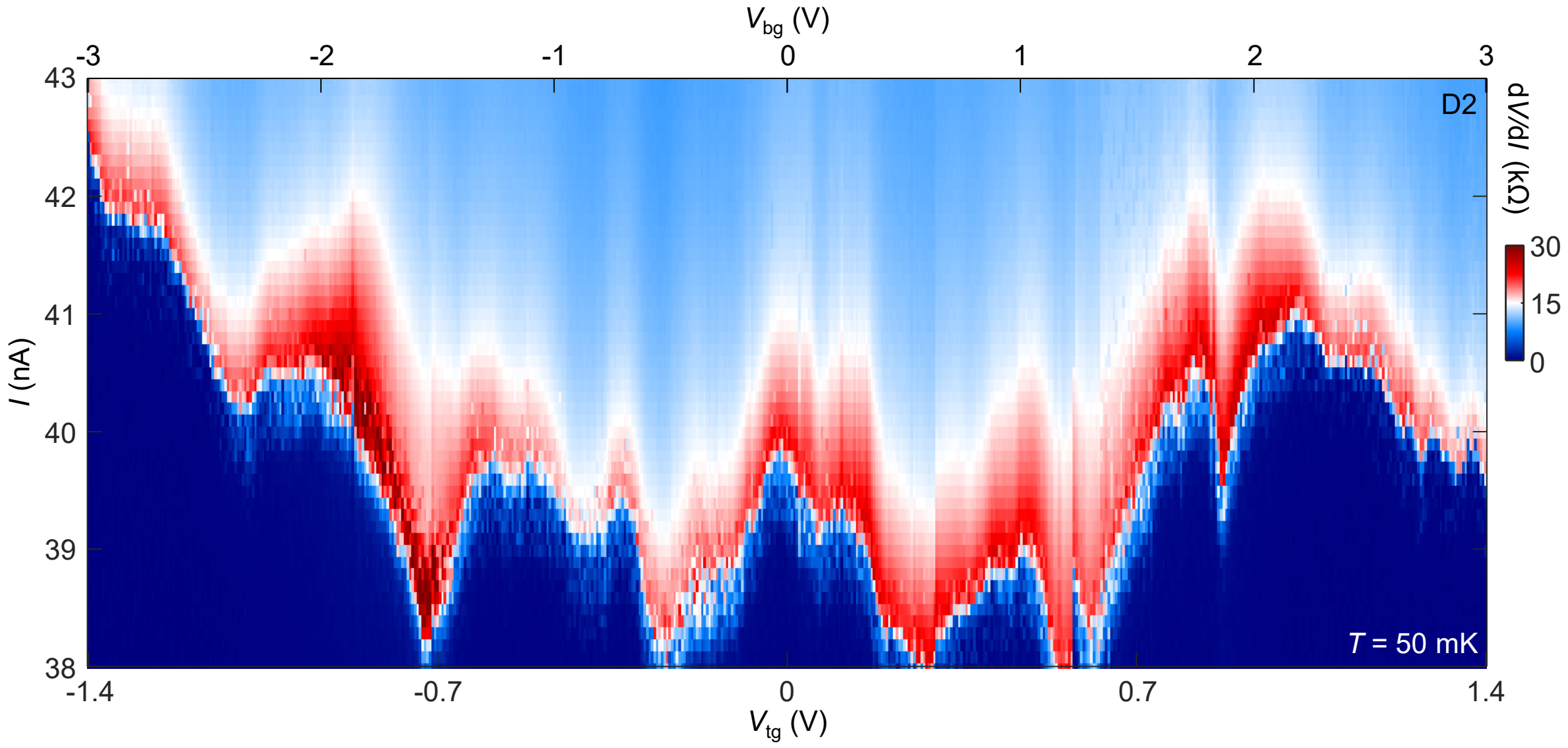

**Fig. S10. Gate modulation of the junction critical current in D2.** Critical current develops weak but interesting modulation in response to gate voltages. Both top ($V_{tg}$) and bottom ($V_{bg}$) gate voltages are simultaneously tuned in this experiment as indicated by bottom and top x-axes. Careful studies of the gate dependent phenomena will be a future focus.